# Metasurface engineering of transient photothermal responses for vibrational imaging at living interfaces

Zhibin Hu[1,2], Tianrun Zhang[1,4], Jiayu Ding[2], Pengcheng Fu[1,4], Ziyang Huang[5], Bo Chen[1,4], Bangjie Song[1,2], Minglu Li[2], Xinyu Deng[1,4], Quan D. Zhou[5*], Delong Zhang[1,4*], Siying Peng[2,3*]

[1]Zhejiang University, Hangzhou, Zhejiang 310027, China.
[2]Department of Materials Science and Engineering, Westlake University, Hangzhou, Zhejiang 310030, China.
[3]Research Center for Industries of the Future, Westlake University, Hangzhou, Zhejiang 310030, China.
[4]Zhejiang Key Laboratory of Micro-nano Quantum Chips and Quantum Control, School of Physics, Zhejiang University, Hangzhou 310027, China.
[5]Institute of Immunology, Department of Surgical Oncology of The First Affiliated Hospital, Zhejiang University School of Medicine, Hangzhou 310058, China.

Corresponding Authors
*Quan D. Zhou, zhouquanzq@zju.edu.cn
*Delong Zhang, dlzhang@zju.edu.cn
*Siying Peng, pengsiying@westlake.edu.cn

## ABSTRACT

Chemical transformations at biological interfaces are central to cellular function, yet the weak vibrational signals associated with subtle molecular changes limit their spatially resolved observation. Here, we present metasurface-driven mid-infrared photothermal imaging (MetaMIP), which uses resonantly engineered metasurfaces to amplify molecular temperature modulation while retaining rapid thermal recovery. Visible-light detection converts this response into submicrometre maps of individual nanoparticles and biological cells. In living cells, MetaMIP resolves spatially distinct chemical changes during enzymatic lipid conversion across the cell membrane. These results demonstrate the potential of engineered photothermal interfaces for mapping chemical composition and tracking its evolution during reactions.

## INTRODUCTION

Understanding molecular interactions and chemical transformations in living systems requires spatially resolved measurements of molecular composition and its evolution.[1] Fluorescence microscopy provides sensitive, high-resolution visualization of labelled targets, but its contrast generally originates from the fluorescent reporter.[2] Vibrational spectroscopy probes intrinsic chemical bonds, providing access to molecular composition and reaction-associated chemical changes.[3] Such information is particularly valuable at biological interfaces, where local interactions and transformations generate spatially heterogeneous molecular distributions.

Infrared (IR) spectroscopy provides molecular vibrational contrast and has become a standard analytical tool across medicine, chemistry, and geology,[4] but faces constraints imposed by the diffraction limit of long-wavelength light and strong water absorption. Mid-infrared photothermal (MIP, also termed O-PTIR) microscopy integrates infrared excitation with visible-light detection of the resulting thermal response, enabling depth-resolved bond-selective imaging with submicrometre spatial resolution.[5-10] Despite these advances, a fundamental limitation arises from the transient photothermal conversion process,[11-13] in which weak absorption within small detection volumes limits the sensitivity.[7,14-16] Improving this sensitivity is needed to distinguish local changes in molecular composition; therefore, it is important for extending vibrational imaging from readily detectable molecular structures to subtle chemical transformations in biological systems.

Resonant metasurfaces offer a route to increasing molecular excitation by concentrating infrared fields near the sample.[17] Recent metasurface-enhanced MIP has mapped the optical hot spots and thermal responses of individual nanoantennas and demonstrated nanomolar molecular detection.[18,19] Extending these capabilities to spatially resolved measurements of dynamic molecular processes requires sufficient photothermal contrast, which represents a relevant design target. Its amplitude and temporal evolution depend on both resonant energy deposition and heat dissipation over an excitation cycle. Heat dissipation influences temperature build-up during excitation but enables recovery after pump termination. A central design objective is therefore to increase the molecular temperature excursion while retaining rapid thermal recovery, connecting resonant optical enhancement to measurable photothermal contrast.

Here, we present MetaMIP, a metasurface-enabled photothermal imaging platform that enhances transient molecular vibrational heating while maintaining rapid post-pulse thermal recovery. We combine measurements of the transient temperature response with coupled electromagnetic-thermal simulations to characterize the temperature rise and relaxation associated with resonant excitation. The enhanced vibrational response supports both scattering- and fluorescence-based detection, enabling measurements across molecular solutions, individual polymer particles, bacteria and living cells. During sphingomyelin hydrolysis in living HeLa cells, MetaMIP distinguishes a broadly distributed decrease in phosphate-associated contrast from a localized increase in a vibrational response consistent with ceramide formation. These measurements demonstrate spatially resolved access to the chemical changes accompanying a biochemical transformation.

## RESULTS

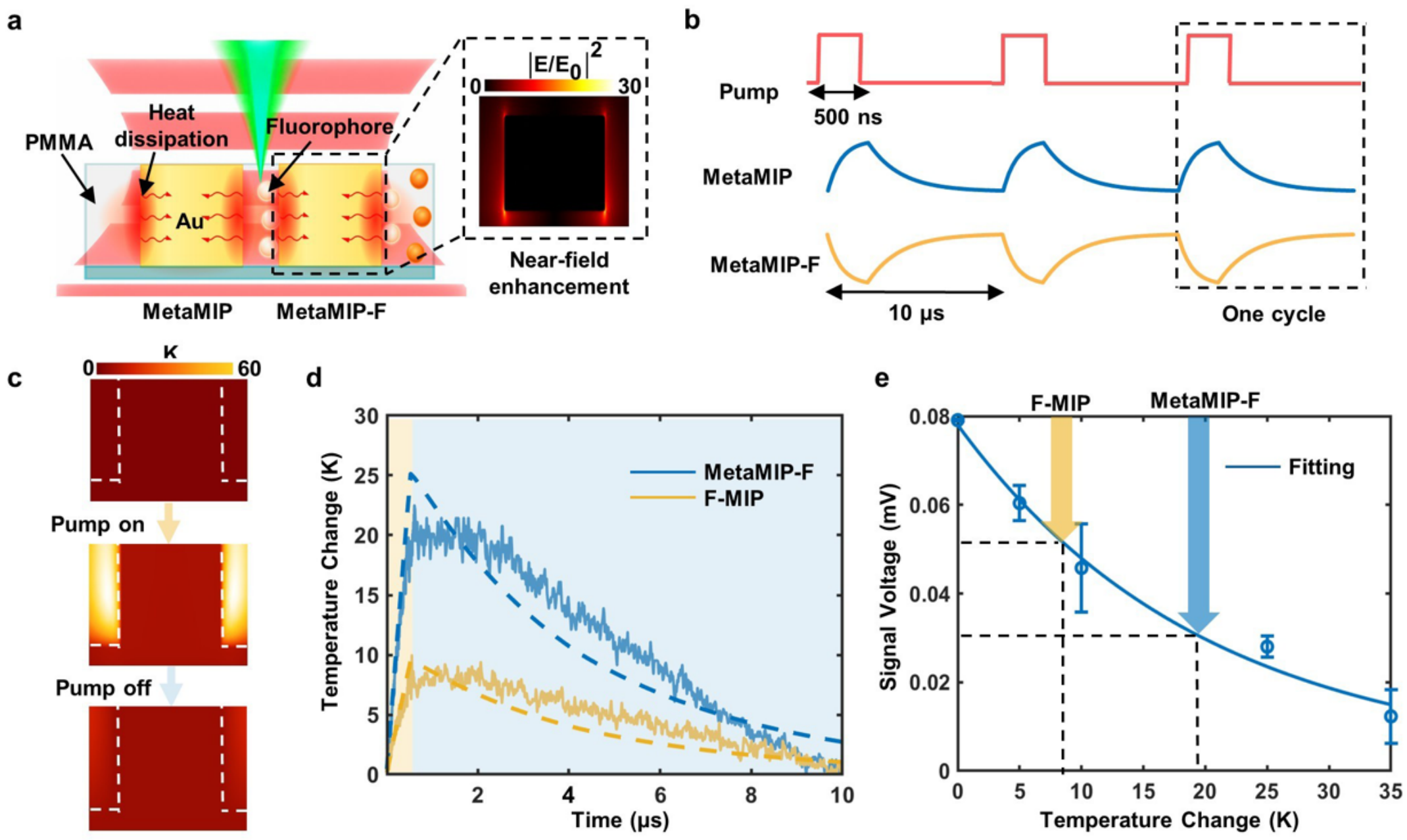

**Fig. 1. Transient photothermal response in MetaMIP.** (a) Schematic of scattering-based MetaMIP and fluorescence-based MetaMIP-F, illustrating resonantly enhanced molecular heating, heat dissipation, and visible-light detection. The simulated near-field intensity enhancement is shown at the metasurface resonance. (b) Excitation and detection timing, showing the 500-ns infrared pump, transient photothermal responses of MetaMIP and MetaMIP-F over successive 10-μs excitation cycles. (c) Simulated temperature distributions during infrared excitation and after pump termination. (d) Experimentally reconstructed (solid) and simulated (dashed) transient temperature responses of PMMA beads with MetaMIP-F (blue) and conventional F-MIP (yellow) over one excitation cycle. (e) Fluorescence-temperature calibration used to reconstruct the experimental temperature transients in d.

MetaMIP couples resonantly enhanced molecular absorption to visible-light detection through the transient thermal response of the sample (Fig. 1a). Infrared fields focused around the metasurface resonators enhance molecular excitation and subsequent local heat generation.[20,21] The resulting temperature modulation is detected through changes in refractive index in scattering-based MetaMIP or, alternatively, through quantum yield modulation in temperature-dependent fluorescence readout MetaMIP (MetaMIP-F). Under our excitation conditions, each 500-ns infrared pulse is followed by a 9.5-μs interval for thermal relaxation, while a continuous-wave visible probe monitors the photothermal response (Fig. 1b).

We designed and fabricated hexagonal metasurfaces with geometries selected to match their resonances to the targeted molecular vibrations. For the design targeting the PMMA carbonyl band at 1730 $cm^{-1}$, electromagnetic simulations showed that coupling between localized plasmon modes and lattice resonances concentrates the infrared field around the resonators, producing a maximum local field intensity enhancement of approximately 40 (Fig. 1a). FTIR characterization confirmed that the fabricated metasurfaces reproduced the designed resonances (Fig. S3). Additional metasurface geometries and their target vibrational bands are detailed in Table S1 and Fig. S2.

To determine how this optical enhancement affects the thermal response, we used the calculated absorption distribution as the heat source for transient thermal simulations under experimentally matched excitation conditions. Heating was initially localized near the resonators and subsequently spread into the surrounding material through thermal diffusion (Fig. 1c). The simulated temperature rise reached approximately 25 K near the metasurface, compared with approximately 10 K on the unpatterned $CaF_2$ substrate (Fig. 1d, dashed lines). Simulations across different resonator geometries showed that the designed radius of 800 nm produced the largest temperature rise (Fig. S5). After pump termination, the temperature relaxed towards the baseline with a characteristic time of approximately 6 μs, limiting thermal accumulation between successive excitation cycles (Fig. S4).

To validate the predicted thermal response, we used 500-nm fluorescent PMMA beads as local temperature probes. Individual beads adjacent to the metasurface or on an unpatterned $CaF_2$ substrate were excited at 1730 $cm^{-1}$, and their fluorescence was recorded synchronously with the infrared pump. An independently measured fluorescence-temperature calibration was used to convert the fluorescence transients into temperature changes (Fig. 1e; Methods and Fig. S8). Beads adjacent to the metasurface reached a peak temperature rise of approximately 20 K, compared with 5-7 K on the unpatterned substrate, followed by relaxation towards the baseline with a characteristic time of approximately 6 μs (Fig. 1d, solid lines). The measurements and simulations thus consistently show a larger transient temperature excursion near the metasurface while retaining submicrosecond thermal recovery.

To compare photothermal enhancement across metasurface architectures, we also examined PMMA on a metasurface supporting a bound state in the continuum (BIC). The normalized MIP image exhibited contrast associated with the patterned structure, but spectra acquired on and outside the BIC metasurface showed no measurable enhancement of the molecular vibrational signal (Fig. S6 and Fig. S7). This comparison highlights the distinction between structural contrast in a photothermal image and enhancement of the molecular vibrational response.

The transient temperature modulation depends on both resonant energy deposition and heat dissipation over the excitation cycle. Heat dissipation reduces the temperature rise during excitation but enables recovery after pump termination. Within a single-exponential thermal model, and neglecting accumulation between pulses, the peak-to-trough temperature modulation at fixed incident pulse energy and effective heat capacity scales as $\Delta T_{mod} \propto G_{eff}\chi_{heat}\chi_{rec}$, where $\chi_{heat} = \frac{1-exp(-t_p/\tau_{th})}{t_p/\tau_{th}}$, $\chi_{rec} = 1 -$ $exp(-t_r/\tau_{th})$ . Here, $G_{eff}$ represents the effective resonant energy-deposition enhancement within the thermally responding sample volume, $\chi_{heat}$ describes thermal retention during the finite pump pulse, and $\chi_{rec}$ describes post-pulse thermal recovery. The parameters $t_p$, $t_r$ and $\tau_{th}$ are the pump duration, interpulse recovery interval, and thermal relaxation time, respectively. Under the present excitation conditions, the 9.5-μs recovery interval substantially exceeds the measured relaxation time, giving $\chi_{rec} \approx 1$. MetaMIP therefore operates with enhanced transient heating and essentially complete thermal recovery between successive pulses. The derivation of the reduced analytical model is provided in Supplementary Note 1.

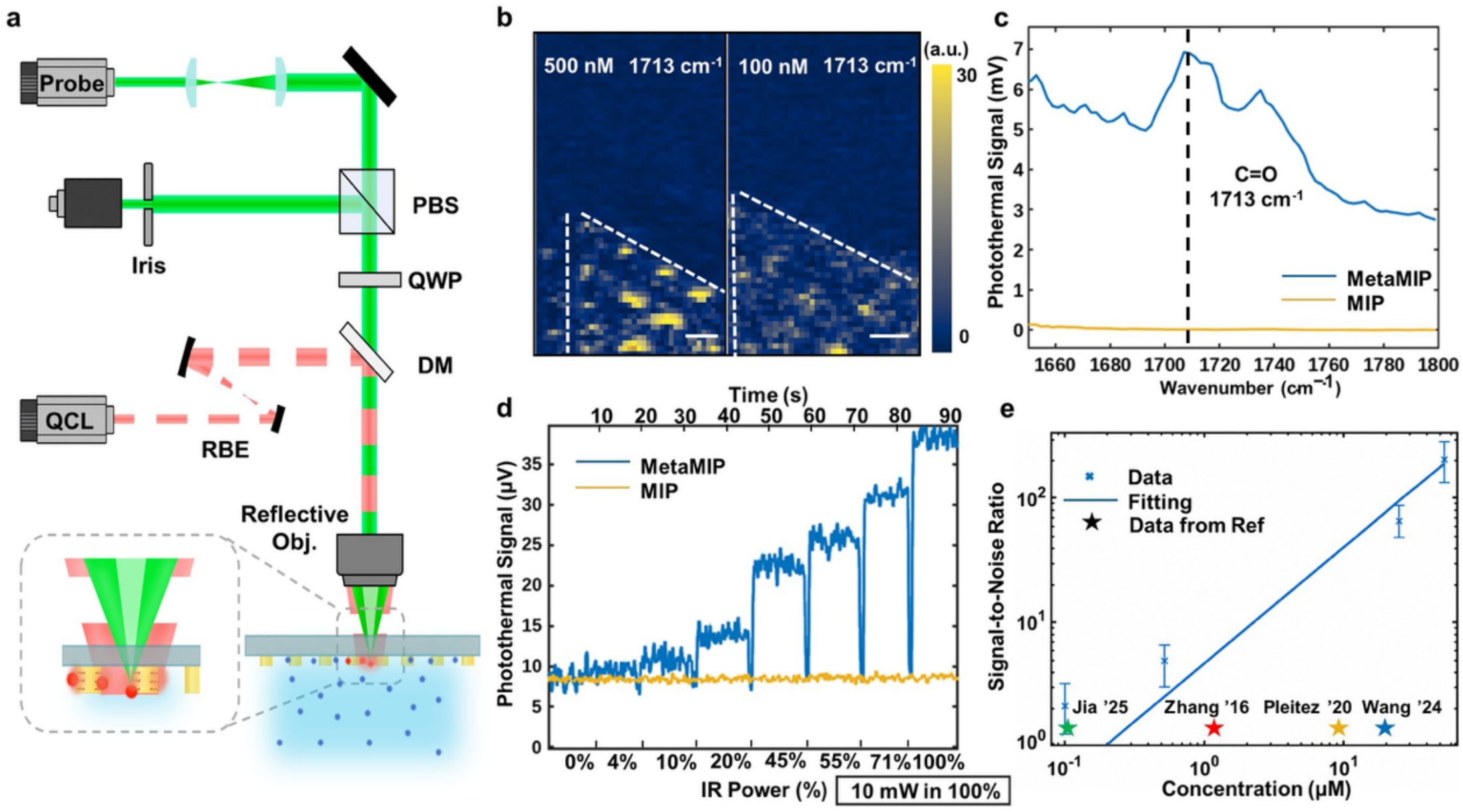

**Fig. 2. Enhanced molecular detection sensitivity of MetaMIP.** (a) Experimental setup for sensitivity measurements. DM, dichroic mirror; BS, beam splitter; BE, beam expander. (b) MetaMIP images of oleic acid solutions at 500 nM (left) and 100 nM (right). Scale bars, 2 μm. (c) MetaMIP (blue) and MIP (yellow) spectra of oleic acid. (d) IR power dependence of signals with (blue) and without (yellow) metasurface at 1713 $cm^{-1}$. (e) Comparison of signal-to-noise ratio versus concentration, showing a 100 nM detection limit. The stars indicate reported detection limits from the literature [5,19,22,23].

The enhanced transient photothermal response is expected to improve molecular detection sensitivity. We therefore quantified the performance of MetaMIP using oleic acid as a model analyte. Oleic acid dissolved in carbon disulfide has a carbonyl stretching vibration (1713 $cm^{-1}$) that matches the resonance of the hexagonal metasurface (Fig. S2). The experimental configuration is illustrated in Fig. 2a. The analyte solution was placed in a quartz well with the metasurface immersed in the solution and examined using a 500 ns, 100 kHz mid-infrared pump together with a continuous-wave 532 nm probe. Photothermal signals acquired with MetaMIP were compared with those with conventional MIP under identical excitation and detection conditions. Control measurements confirmed that the quartz well did not contribute appreciably to the measured photothermal signal (Fig. S24).

MetaMIP images acquired from oleic acid solutions at concentrations of 500 nM and 100 nM are shown in Fig. 2b. At both concentrations, photothermal contrast was clearly observed with MetaMIP, whereas the MIP signals remained at the noise level. The corresponding photothermal spectra (Fig. 2c) show a distinct carbonyl vibration at 1713 $cm^{-1}$ with MetaMIP, while no detectable peak was observed with MIP. An additional peak at 1733 $cm^{-1}$ was observed with MetaMIP. Control measurements using pure carbon disulfide confirmed that this feature originated from the solvent and was independent of the carbonyl vibration of oleic acid (Fig. S10).

We next examined the dependence of the photothermal signal on mid-infrared excitation power (Fig. 2d). As the pump power increased from 0 to 10 mW, the MetaMIP signal increased monotonically, whereas the conventional MIP signal remained close to the noise floor over the same power range. The average photothermal signal extracted at each power level exhibited an approximately linear dependence on pump power. To exclude contributions from laser fluctuations or cumulative thermal effects, the signal was continuously monitored for 5 min at 10 mW. The measured signal fluctuated around 40 μV without measurable drift (Fig. S10), confirming the stability of the measurement.

The molecular detection limit of MetaMIP was quantified using oleic acid solutions spanning concentrations from 50 μM to 100 nM (Fig. 2e). Photothermal signals were acquired at 1713 $cm^{-1}$ under identical excitation conditions, and the signal-to-noise ratio (SNR) was determined for each concentration. The SNR was calculated as the mean photothermal signal within the selected region of interest divided by the standard deviation of the probe signal. The SNR decreased with decreasing analyte concentration and reached approximately 3 at 100 nM, which was used as the experimental detection limit. MetaMIP therefore achieved a detection limit of approximately 100 nM for oleic acid under the present experimental conditions. This represents an approximately 100-fold improvement over the micromolar sensitivity typically reported for conventional MIP and is compared with representative vibrational imaging approaches in Table S2[5, 19, 22, 23]. Additional measurements characterized the wavelength tunability, spectral accuracy, pump-power and polarization dependence, and off-resonance response of MetaMIP (Figs. S11-S17).

The enhanced photothermal sensitivity of MetaMIP enables vibrational imaging of nanoscale objects that produce signals below the detection capability of conventional MIP under the same measurement conditions. To evaluate the imaging capability of MetaMIP, PMMA beads with a diameter of 100 nm were deposited onto $CaF_2$ substrates containing both metasurface and unpatterned regions. The samples were imaged using a scanning photothermal microscope with a step size of 100 nm. The photothermal images were obtained by normalizing the raw photothermal signal to the corresponding DC image (Methods and Fig. S22). The mid-infrared pump was tuned to the carbonyl stretching vibration of PMMA at 1730 $cm^{-1}$. MetaMIP clearly detected individual 100 nm PMMA beads with high photothermal contrast (Fig. 3a,b). In contrast, conventional MIP produced substantially weaker signals that were difficult to distinguish from the background (Fig. 3c,d). The corresponding DC images identify the locations of the individual PMMA beads and confirm that the observed photothermal signals originate from the target objects.

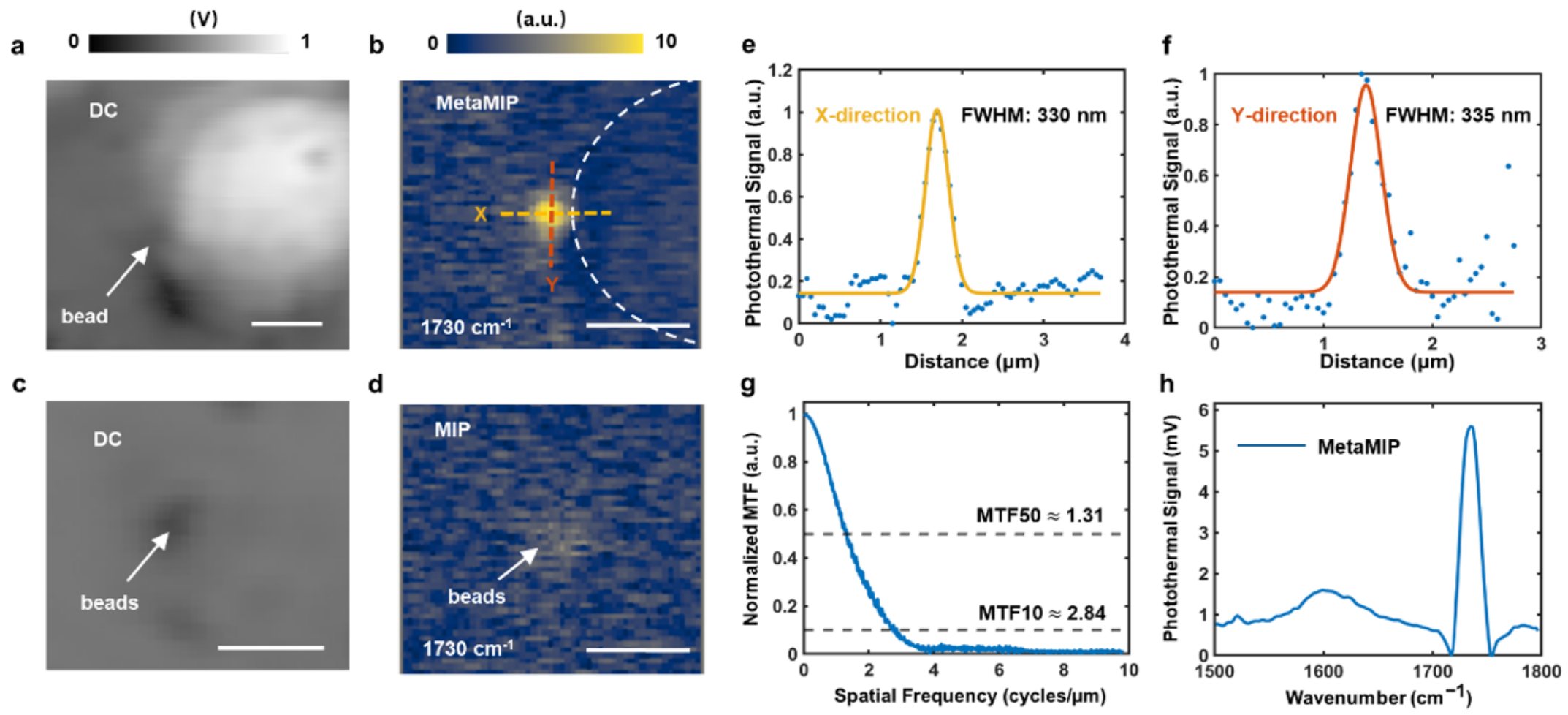

**Fig. 3. Far-field vibrational imaging of 100 nm PMMA beads using MetaMIP.** (a, b) DC and MetaMIP images of a single 100-nm PMMA bead on the metasurface. (c, d) Corresponding DC and MIP images acquired on an unpatterned substrate. Arrows indicate individual PMMA beads, and dashed circles outline the metasurface region. (e, f) Cross-sectional photothermal intensity profiles along the x and y directions of the MetaMIP image, yielding full widths at half maximum (FWHM) of 330 nm and 335 nm, respectively. (g) Normalized modulation transfer function (MTF) derived from the reconstructed MetaMIP image. (h) Photothermal spectrum acquired from a single 100-nm PMMA bead using MetaMIP. Scale bars, 1 μm.

To characterize the spatial imaging performance of MetaMIP, an isolated 100 nm PMMA bead located within a near-field hotspot along the edge of the metasurface structure was selected for analysis. Cross-sectional photothermal intensity profiles extracted along the x and y directions through the signal maximum of an isolated PMMA bead yielded full widths at half maximum (FWHM) of 330 nm and 335 nm, respectively (Fig. 3e,f). The modulation transfer function (MTF) was estimated from the MetaMIP image of a single PMMA bead by calculating the magnitude of its two-dimensional Fourier transform, normalizing it to the zero-frequency component, and performing a 360° radial average in the spatial-frequency domain (Fig. 3g). The resulting MTF complements the real-space FWHM measurements by characterizing the spatial-frequency response of MetaMIP.

To establish that the signal detected from the 100 nm PMMA beads retains molecular vibrational specificity, a photothermal spectrum was acquired from an individual PMMA bead near the metasurface (Fig. 3h). A pronounced peak was observed at 1730 $cm^{-1}$, corresponding to the characteristic C=O stretching vibration of PMMA. The coincidence of the photothermal feature with the particle position in the DC image, together with the characteristic vibrational peak, confirms that the detected contrast originates from PMMA. These measurements demonstrate far-field vibrational imaging of individual 100 nm PMMA beads while preserving molecular specificity. We further evaluated the MetaMIP-F system using 100 nm PMMA beads. Individual 100 nm PMMA beads were also detected by MetaMIP-F (Fig. S9), demonstrating the applicability of fluorescence-based detection to nanoscale objects.

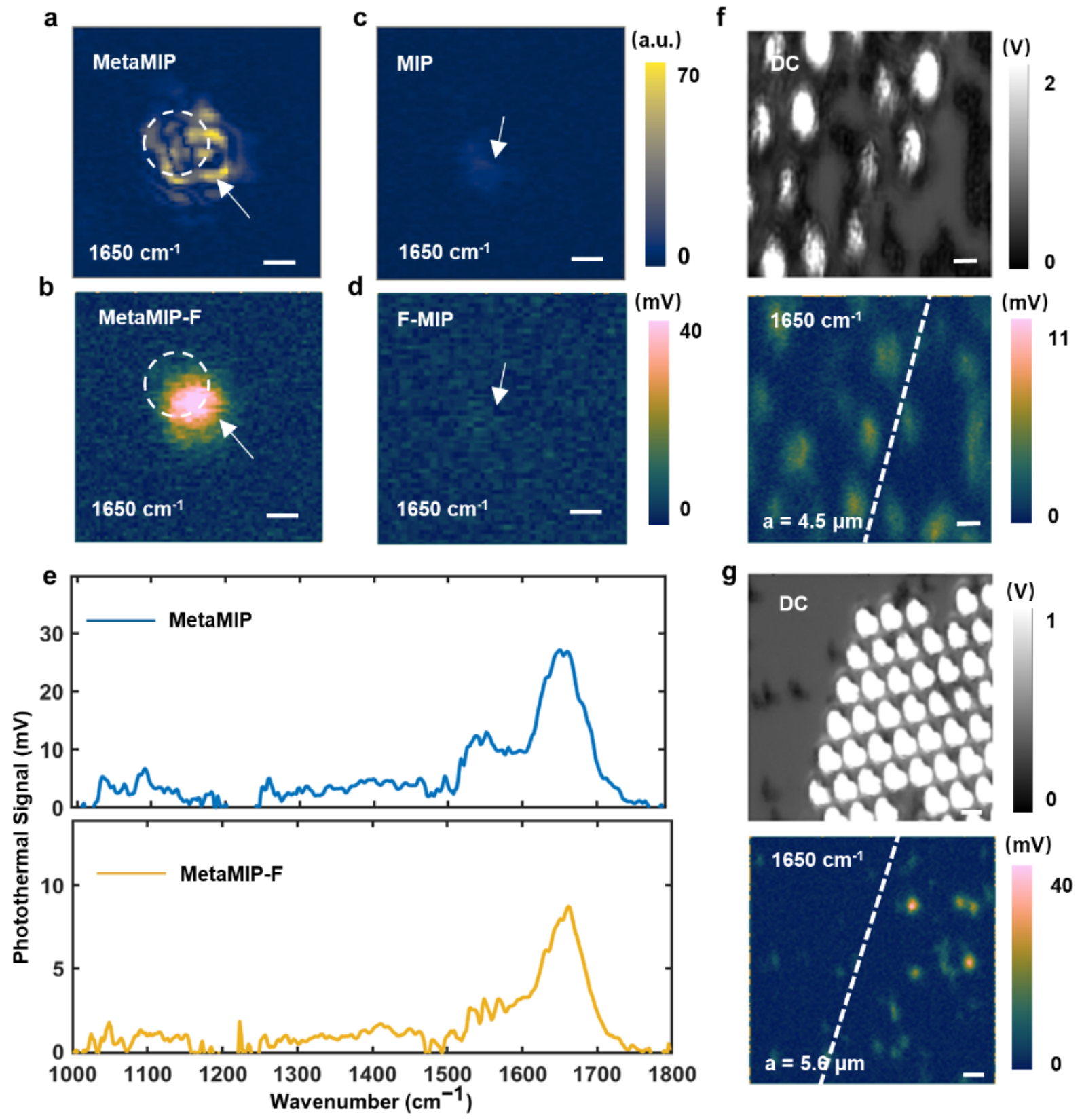

**Fig. 4. Bond-selective vibrational imaging of individual bacterium using MetaMIP.** (a, b) Scattering- and fluorescence-based MetaMIP images of individual *S. aureus* cells. (c, d) Corresponding MIP and F-MIP images. Arrows indicate individual *S. aureus* cells and dashed circles outline the metasurface. (e) Photothermal spectra acquired using scattering-based MetaMIP (top) and fluorescence-based MetaMIP-F (bottom). (f, g) Fluorescence-based MetaMIP images of *S. aureus* acquired at 1650 $cm^{-1}$ using metasurfaces resonant at 1730 $cm^{-1}$ (4.5 μm periodicity) and 1650 $cm^{-1}$ (5.0 μm periodicity), respectively. The dashed line indicates the boundary of the metasurface region. Scale bars, 1 μm.

To evaluate the capability of MetaMIP for biological vibrational imaging, individual inactivated *Staphylococcus aureus* (*S. aureus*) cells were imaged using both scattering- and fluorescence-based photothermal detection.[24-26] *S. aureus* is a clinically important bacterial pathogen with a characteristic diameter of 0.5-1.5 μm, providing a representative model for evaluating chemically selective imaging of a single bacterium.

With the mid-infrared pump tuned to 1650 $cm^{-1}$, scattering-based MetaMIP clearly detected individual bacterial cell with high photothermal contrast (Fig. 4a), whereas conventional MIP acquired under the same excitation conditions produced substantially weaker signals that were difficult to distinguish from the background (Fig. 4c). A similar enhancement was observed using fluorescence-based detection: MetaMIP-F resolves individual bacterium with high contrast (Fig. 4b), while conventional F-MIP fails to distinguish the cells from the background (Fig. 4d). The detected bacterial features had an apparent size of approximately 1 μm, consistent with the dimensions of individual *S. aureus* cell.

To verify the vibrational origin of the detected contrast, photothermal spectra were acquired from the bacterial cells using both detection modes (Fig. 4e). Both scattering- and fluorescence- based measurements exhibited a pronounced peak at 1650 $cm^{-1}$, corresponding to the amide I vibration of *S. aureus*. The agreement between the two independently detected spectra confirms that the enhanced contrast originates from the targeted molecular vibration rather than from nonspecific optical contrast.

The bond selectivity of MetaMIP was further examined by varying the resonance of the metasurface. When *S. aureus* cells were deposited onto a metasurface resonant with the carbonyl vibration at 1730 $cm^{-1}$ (4.5 μm periodicity), excitation at 1650 $cm^{-1}$ produced no appreciable enhancement of the bacterial photothermal signal within the metasurface region (Fig. 4f). In contrast, a metasurface resonant with the amide I vibration at 1650 $cm^{-1}$ (5.0 μm periodicity) produced a pronounced enhancement from bacteria located within the metasurface region (Fig. 4g). The enhancement therefore occurs only when the metasurface resonance is matched to the targeted molecular vibration, confirming the bond-selective response of the MetaMIP platform.

To investigate whether MetaMIP enables bond-selective visualization of dynamic membrane chemistry at the living cell-substrate interface, we monitored the hydrolysis of sphingomyelin (SM) in the plasma membrane of HeLa cells. In the presence of water, sphingomyelinase from *Bacillus cereus* (bSMase) hydrolyses SM into ceramide (Cer), which remains in the plasma membrane, and water-soluble phosphocholine (PCho), which is released into the extracellular medium (Fig. 5a). Ceramide generated through SM hydrolysis can alter membrane organization, making the spatial evolution of this reaction biologically relevant[27]. Consequently, the phosphocholine headgroup of SM, which exhibits a characteristic symmetric $PO2^{-}$ stretching vibration near 1085 $cm^{-1}$, is progressively depleted during hydrolysis, whereas Cer lacks this phosphate-containing group. HeLa cells were cultured directly on a MetaMIP substrate resonant at 1085 $cm^{-1}$, enabling the phosphate-associated signal from the basal plasma membrane to be monitored throughout the reaction.

Photothermal spectra acquired from the basal plasma membrane showed a pronounced phosphate-associated band at 1085 $cm^{-1}$ with MetaMIP, which was substantially weaker with conventional MIP (Fig. 5b). To follow the temporal evolution of the reaction, different HeLa cells were imaged by MetaMIP at 0, 10, and 20 min after the initiation of bSMase treatment (Fig. 5f-h). Before hydrolysis, strong photothermal contrast was observed from the basal plasma membrane at 1085 $cm^{-1}$, delineating the region where the membrane was in close proximity to the metasurface. Following bSMase treatment, the phosphate-associated signal progressively decreased with reaction time. After 10 min, the signal intensity was substantially reduced, and after 20 min, only weak residual contrast remained. The time-dependent decrease in the 1085

cm⁻¹ signal visualizes the progressive loss of phosphate-associated membrane contrast during SM hydrolysis.

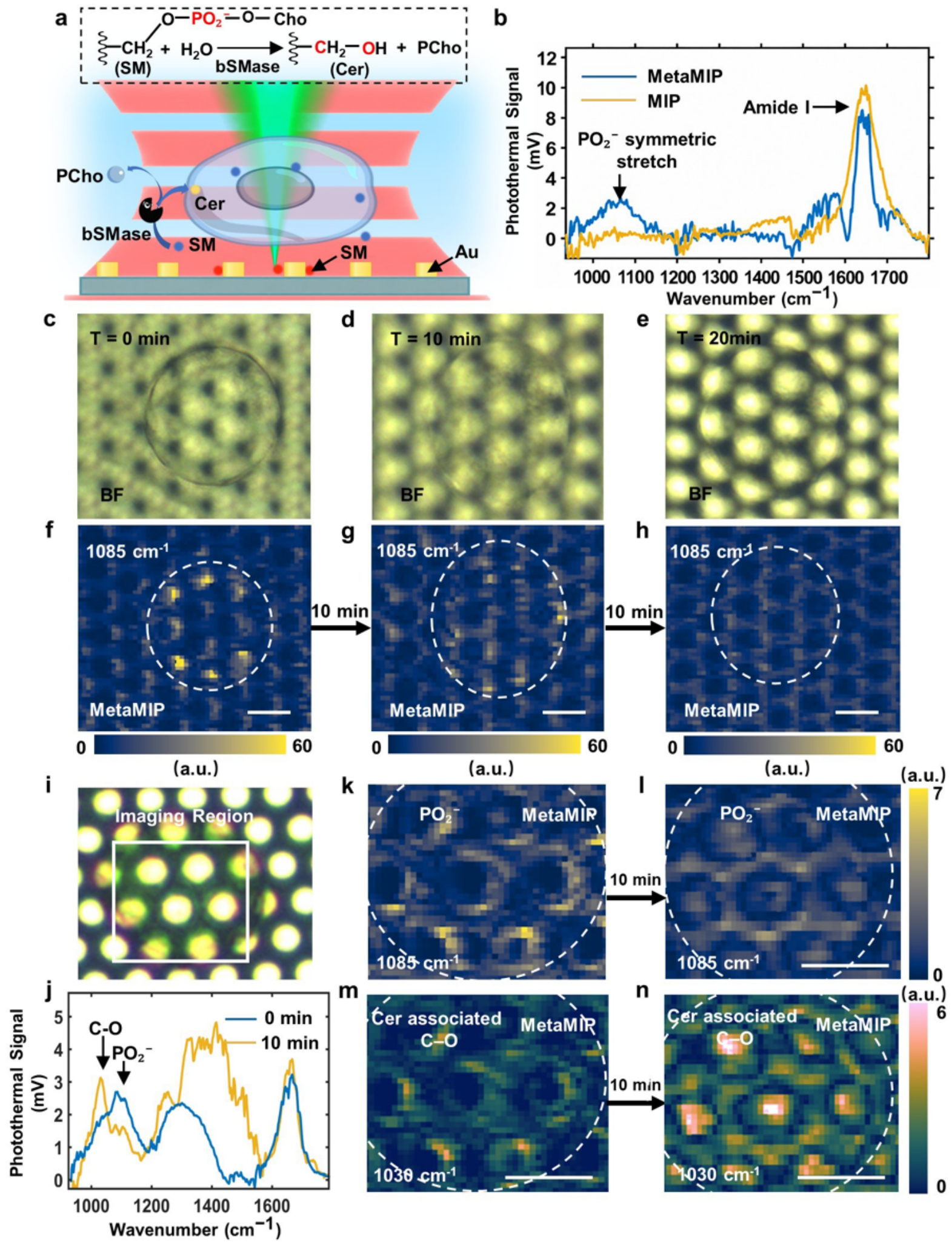

**Fig. 5. Bond-selective vibrational imaging of membrane lipid reactions in living cells using MetaMIP.** (a) Schematic of the hydrolysis of sphingomyelin (SM) into ceramide (Cer) and phosphocholine (PCho), catalysed by sphingomyelinase from *Bacillus cereus* (bSMase). (b) Photothermal spectra acquired from the basal plasma membranes of HeLa cells using MetaMIP and conventional MIP. (c-e) Optical micrographs of different HeLa cells cultured on the MetaMIP substrate at 0, 10 and 20 min after the initiation of SM hydrolysis, respectively. (f-h) Corresponding MetaMIP images acquired at 1085 cm⁻¹. (i) Optical micrograph of a representative HeLa cell used to track membrane-associated vibrational changes during SM hydrolysis. (j) MetaMIP spectra acquired from the same region of this cell before and 10 min after the onset of hydrolysis. (k, l) MetaMIP images acquired from the same cell region at 1085 cm⁻¹ before and 10 min after the onset of hydrolysis, respectively, showing the evolution of phosphate-associated membrane contrast. (m, n) Corresponding MetaMIP images acquired from the same cell region at 1030 cm⁻¹, showing the evolution of C–O-associated membrane contrast. Scale bars, 5 μm.

Under identical imaging conditions, conventional MIP produced substantially weaker membrane contrast at 1085 cm⁻¹ and did not clearly distinguish the signal changes at successive stages of SM hydrolysis (Fig. S18a-f). In comparison, MetaMIP provided sufficient vibrational contrast to follow the time-dependent reduction of the phosphate-associated membrane signal, enabling bond-selective visualization of membrane lipid chemistry at the living cell-substrate interface.

We next examined the spatial evolution of membrane-associated vibrational signatures during SM hydrolysis by imaging the same region of an individual living HeLa cell before and after the reaction. Fig. 5i shows an optical micrograph of the cell selected for observation. Fig. 5j compares MetaMIP spectra acquired from the same cell at the same region before and 10 min after the onset of hydrolysis. Before the reaction, the spectrum (blue curve) exhibited a pronounced phosphate-associated band at approximately 1085 $cm^{-1}$, consistent with the phosphocholine headgroup of SM. After 10 min of hydrolysis (yellow curve), the 1085 $cm^{-1}$ band remained detectable but decreased markedly in intensity, whereas the band around 1030 $cm^{-1}$ was substantially enhanced. Notably, a spectral contribution around 1030 $cm^{-1}$ was already present before hydrolysis,[28-30] although it appeared only as a weak shoulder because of the much stronger neighboring SM-associated band at 1085 $cm^{-1}$. The pronounced enhancement of the 1030 $cm^{-1}$ band following hydrolysis indicates an increased contribution from C–O vibrations and is consistent with the formation and accumulation of Cer. An increase in the spectral response around 1400 $cm^{-1}$ was also observed after hydrolysis, potentially reflecting overlapping contributions from $CH_3$ deformation, $CH_2$ bending and symmetric carboxylate stretching vibrations.

To directly visualize this chemical transformation, MetaMIP imaging was performed at 1085 and 1030 $cm^{-1}$ before and after hydrolysis. Fig. 5k,l show the spatial distribution of the phosphate-associated membrane signal before and 10 min after the reaction, respectively. A pronounced reduction in the phosphate-associated signal was observed after hydrolysis, consistent with the trend observed in Fig. 5f,g. Interestingly, the MetaMIP image acquired at 1030 $cm^{-1}$ before hydrolysis (Fig. 5m) exhibited a hotspot distribution closely resembling that observed at 1085 $cm^{-1}$ in Fig. 5k, suggesting that the initial signal in this spectral region was predominantly associated with the pre-existing membrane components, including SM. In contrast, after 10 min of hydrolysis, the signal at 1030 $cm^{-1}$ increased substantially (Fig. 5n), consistent with Cer formation. Notably, following hydrolysis, the enhanced 1030 $cm^{-1}$ signal became spatially concentrated in the central cell-metasurface contact region, whereas the phosphate-associated signal at 1085 $cm^{-1}$ decreased more broadly across the membrane. This spatial evolution was reproduced in additional HeLa cells (Fig. S18g,h), revealing spatially heterogeneous membrane chemistry during SM hydrolysis. Additional MetaMIP imaging at 1730 $cm^{-1}$ resolved carbonyl-associated lipid contrast in living HeLa cells (Fig. S19). Together, these results show that MetaMIP resolves the coupled temporal and spatial evolution of membrane-associated vibrational signatures during a biochemical reaction in living cells.

## DISCUSSION

MetaMIP integrates resonant infrared excitation with visible-light photothermal detection to enhance bond-selective imaging at biological interfaces. The design objective is to increase the transient temperature excursion while preserving thermal recovery between excitation cycles. Experimental reconstruction of the transient temperature evolution, together with coupled electromagnetic-thermal simulations, shows that MetaMIP achieves a substantially larger temperature modulation while retaining rapid post-pulse thermal relaxation. These findings identify the transient temperature modulation as a useful criterion for evaluating metasurface performance, alongside electromagnetic field enhancement.

The enhanced transient photothermal response expands beyond detecting weaker vibrational signals to resolving spatial differences in reaction-associated chemical changes. During sphingomyelin hydrolysis in living cells, the broadly distributed decrease in phosphate-associated contrast and the more localized increase in a response consistent with ceramide formation reveal distinct spatial patterns within the overall spectral change. Such measurements distinguish where different vibrational signatures evolve, information that is obscured by spatial averaging. This spatial distinction is particularly relevant to sphingomyelin metabolism, in which ceramide formation can reorganize membrane structure and contribute to the spatial organization of cellular signaling[31,32]. Demonstrations across molecular samples, individual nanoparticles, bacteria and living cells establish the applicability of MetaMIP to different sample classes. Its implementation with both scattering and fluorescence readout further shows that enhanced vibrational excitation can be detected through different optical observables, with chemical selectivity supplied by infrared absorption.

Quantitative interpretation nevertheless requires consideration of both molecular composition and coupling to the metasurface. The confined enhancement preferentially probes material near the resonators, making the signal sensitive to sample position and, in cellular measurements, membrane-substrate separation. Reaction-associated intensity changes must therefore be distinguished from changes in contact geometry. Moreover, measurements across multiple bands and enhanced spectral selectivity will strengthen molecular assignments and quantitative analysis. Future integration of multiresonant or tunable metasurfaces with calibrated, faster photothermal acquisition could extend these measurements to several reaction-associated signatures within the same field of view. This would enable spatially resolved studies of chemical transformations at biological interfaces, linking changes in molecular composition to their local environment.

## MATERIALS AND METHODS

### Metasurface fabrication

Double-side polished calcium fluoride ($CaF_2$) substrates (1 × 1 cm) were sequentially cleaned with acetone, isopropanol (IPA), and deionized water, followed by baking at 105 °C for 10 min. A 2 μm-thick photoresist layer (ARP-5350, Allresist) was prepared by spin coating at 1000 rpm for 45 s and soft-baked at 105 °C for 3 min.

Metasurface patterns were defined using a maskless photolithography system (Microwriter ML3, Durham Magneto Optics). Exposure doses of 220, 190, and 160 mJ $cm^{-2}$ were used for metasurfaces designed for the carbonyl (1730 $cm^{-1}$), amide II (1550 $cm^{-1}$), and phosphate (1080 $cm^{-1}$) vibrational bands, respectively. After exposure, the samples were developed in AR300-26 developer diluted with deionized water (1:7, volume ratio) for 75 s and rinsed with deionized water.

A 5 nm titanium adhesion layer followed by a 500 nm gold layer was deposited using a high-vacuum thermal evaporator (ei-5z, ULVAC). Lift-off was performed in acetone, followed by rinsing with IPA and drying under nitrogen. The thickness of the deposited metal film was verified using a surface profilometer (Fig. S20). The geometrical parameters of the fabricated Mie resonators and hexagonal metasurface are summarized in Fig. S1, Fig. S21, and Table S1.

**Photothermal imaging instrumentation**

Photothermal imaging was performed using a commercial mid-infrared photothermal microscope (mIRage-R, Photothermal Spectroscopy Corp.). The system employed a reflective objective (PIKE, 40×, NA = 0.78) for both infrared excitation and visible-light detection. Mid-infrared excitation was provided by a tunable quantum cascade laser (MIRcat, Daylight Solutions) operating at a repetition rate of 100 kHz with a pulse width of 500 ns and a duty cycle of 5%. Unless otherwise specified, the average infrared power was 10 mW, corresponding to an average irradiance of approximately 5.7 kW $cm^{-2}$ with a focal spot diameter of ~15 μm at 1730 $cm^{-1}$. A continuous-wave 532 nm laser served as the probe beam and was operated at 3% of the maximum output power.

For scattering-based MetaMIP measurements, the reflected probe beam was detected using the built-in balanced photodetector of the microscope, and the photothermal signal was demodulated by a lock-in amplifier synchronized to the infrared pump modulation. For fluorescence-based MetaMIP-F measurements, fluorescence emission was collected in the backward direction using a photomultiplier tube (7422-40, Hamamatsu) positioned behind a 550 nm long-pass filter (550LP, Thorlabs), while the remaining optical configuration was identical to that used for scattering-based measurements.

Before image acquisition, the infrared pump beam was fixed, and the visible probe beam was aligned by scanning across a standard low-emissivity glass substrate to maximize the demodulated photothermal signal. Photothermal images were acquired by raster scanning the sample with a motorized translation stage. Unless otherwise specified, images were recorded with a pixel dwell time of 10 μs over a field of view of 50 × 50 $\mu m^2$.

For bacterial imaging, inactivated PI-labelled *S. aureus* cells were measured on metasurfaces resonant with the amide I vibration at 1650 $cm^{-1}$ using both scattering-based MetaMIP and fluorescence-based MetaMIP-F.

**Fluorescence-based transient temperature reconstruction**

Transient temperature measurements were performed using 500 nm Rhodamine B-labelled PMMA beads as fluorescent temperature probes. The fluorescence detection path of the MetaMIP system was modified by replacing the original photodiode with a photomultiplier tube (7422-40, Hamamatsu). A 550 nm long-pass filter (550LP, Thorlabs) was positioned before the detector to reject the 532 nm probe beam while transmitting the fluorescence emission. The photomultiplier output was recorded using a digital oscilloscope, while the mid-infrared pulse monitored by the mercury cadmium telluride (MCT) detector was simultaneously recorded through a second oscilloscope channel for synchronization.

The 532 nm probe beam and a 1730 $cm^{-1}$ mid-infrared pump pulse (500 ns pulse width) were focused onto individual PMMA beads located either adjacent to the metasurface or on an unpatterned $CaF_2$ substrate. Unless otherwise stated, the mid-infrared power was maintained at approximately 7 mW to avoid thermal deformation of the PMMA beads while maintaining sufficient signal-to-noise ratio.

Raw fluorescence traces were corrected by subtracting the dark voltage of the photomultiplier tube and processed using matched filtering to recover the transient fluorescence response.[33] A fluorescence-temperature calibration was established using PMMA beads placed on a temperature-controlled heating stage by measuring the fluorescence intensity at room temperature and temperature increases of 5, 10, 25, and 35 K after thermal equilibration for 3 min at each temperature. Three measurements were acquired under each condition and averaged to generate the calibration curve. The fluorescence-temperature relationship was fitted using an empirical exponential decay function, which was subsequently used to convert transient fluorescence signals into transient temperature evolution.

**Sample preparation**

Oleic acid samples

Oleic acid (Sigma-Aldrich) was serially diluted in carbon disulfide ($CS_2$) to the desired concentrations immediately before measurement. For sensitivity measurements, 100 μL of solution was introduced into a quartz well (0.8 cm × 0.8 cm cross-section, 1 cm depth) placed directly on the metasurface substrate, with the metasurface serving as the bottom window of the well. Control measurements were performed on bare $CaF_2$ substrates under identical conditions.

PMMA bead samples

Two types of PMMA beads were used in this work. Rhodamine B-labelled PMMA beads with a diameter of 500 nm were employed as fluorescent temperature probes for transient temperature reconstruction, whereas unlabelled PMMA beads with a diameter of 100 nm were used for nanoscale photothermal imaging and spectral measurements. The bead suspensions were deposited onto the metasurface substrates and allowed to dry under ambient conditions before imaging.

*Staphylococcus aureus*

Glycerol stock of *Staphylococcus aureus* (ATCC 29213) was inoculated into sterile tryptone soy broth media and cultured at 37 °C, 220 rpm overnight. Live *S.aureus* cell suspension was centrifuged and washed twice with phosphate-buffered saline (PBS), followed by a heat-inactivation procedure (60 °C, 2 h). The inactivated bacterial suspension was stained with propidium iodide (PI) at 1 μg $mL^{-1}$ working concentration for 10 min in the dark at room temperature. The bacterial suspension was centrifuged twice at 4000 rpm for 5 min, with washing in PBS between centrifugation steps. After the final wash, the cells were resuspended in deionized water, deposited onto the metasurface substrates, and dried at 50 °C for 10 min before imaging. The culture, storage, and heat inactivation steps were all carried out in a specialized BSL-2 laboratory to ensure biosafety.

HeLa cell culture and sphingomyelinase treatment

HeLa cells were cultured in Dulbecco's modified Eagle's medium (DMEM) supplemented with 10% fetal bovine serum and 1% penicillin–streptomycin at 37 °C in a humidified atmosphere containing 5% $CO_2$. Before cell seeding, metasurface-patterned $CaF_2$ substrates were sterilized in 75% ethanol for 1 h, rinsed with deionized water, and transferred to 12 well culture plates. Cells were seeded at a density of $1.5 \times 10^5$ cells $mL^{-1}$ and cultured overnight to allow complete attachment. Before imaging, the substrates were washed twice with PBS and transferred to serum-free DMEM containing 1% penicillin–streptomycin. Sphingomyelin hydrolysis was initiated by adding sphingomyelinase from *Bacillus cereus* (bSMase) to a final concentration of 25 mU $mL^{-1}$. Independent groups of live HeLa cells were then imaged after 0, 10, or 20 min of bSMase treatment, respectively, under otherwise identical imaging conditions.

**Image processing and quantitative analysis**

Raw photothermal images were normalized to the corresponding direct-current (DC) reflection images to compensate for spatial variations in probe intensity and sample reflectivity. The normalized photothermal signal was calculated by dividing the lock-in photothermal image by the corresponding DC reflection image on a pixel-by-pixel basis. This normalization reduced background variations arising from optical reflection and illumination nonuniformity (Fig. S22). Meanwhile, Fig. S23 shows that the bare metasurface without any sample produces no pronounced photothermal signal, excluding a significant contribution from the gold metasurface itself to the experimentally measured response.

For fluorescence-based transient temperature measurements, the raw photomultiplier voltage traces were corrected by subtracting the mean dark voltage and subsequently processed using matched filtering to recover the transient fluorescence response to each mid-infrared pulse. The processed fluorescence signals were converted into transient temperature evolution using the fluorescence-temperature calibration described above. Photothermal spectra were obtained by extracting the average photothermal intensity from regions of interest corresponding to the selected samples or metasurface hotspots while scanning the mid-infrared excitation wavelength across the target vibrational bands. Unless otherwise specified, spectra shown in the manuscript were normalized to facilitate comparison between different samples and measurement conditions.

Spatial resolution was quantified from photothermal images of individual 100 nm PMMA beads. Cross-sectional intensity profiles were extracted along the horizontal and vertical directions, and the full width at half maximum (FWHM) was obtained by profile fitting. The MTF was estimated from the MetaMIP image of an isolated 100-nm PMMA bead. A two-dimensional Fourier transform was applied to the particle image, and the magnitude of the Fourier spectrum was calculated and shifted such that the zero spatial frequency was located at the center. The Fourier magnitude was normalized to its zero-frequency value. A 360° radial average about the frequency origin was subsequently performed to obtain the one-dimensional MTF as a function of spatial frequency.

**Electromagnetic and thermal simulations**

Coupled electromagnetic and transient thermal simulations were performed using COMSOL Multiphysics. Electromagnetic simulations were carried out to obtain the local electric-field distribution of the metasurface at the target vibrational frequencies. The geometrical parameters of the metasurface were identical to those used experimentally.

The calculated electromagnetic field distribution was subsequently used to determine the spatial distribution of optical absorption within the PMMA layer, which served as the heat source for transient thermal simulations. Heat transfer was simulated under excitation by a mid-infrared pulse using experimentally matched excitation conditions. The temporal evolution of the temperature field was calculated by solving the transient heat conduction equation.

**Acknowledgements**
This work was supported by the National Natural Science Foundation of China (12674398), National Key Research and Development Program of China (2024YFA1408900), National Natural Science Foundation of China (32171275 and 32371330), SOE-DSPF of Westlake University, Research Center for Industries of the Future, Westlake Education Foundation, Russian Science Foundation (26-79-31011), the Zhejiang Provincial Natural Science Foundation of China (LQK26H100001) and Fundamental Research Funds for the Central Universities (226-2024-00161). We thank Zhong Chen and Xue Lou for technical assistance, and Weilin Hu and Chunlian Li for their assistance in S. aureus experiments. *Staphylococcus aureus* (ATCC 29213) is a kind gift from Fei Xu at ZJU. We thank the Instrumentation and Service Center for Molecular Sciences and Micro/Nano Fabrication at Westlake University for the facility support and technical assistance.